\documentclass[acus]{JAC2003}

\usepackage{graphicx}
\usepackage{booktabs}

\usepackage{mathptmx}
\newcommand{\q}[2]{\ensuremath{#1\ \mathrm{#2}}}

\begin{document}

\title{BEAM-BEAM COMPENSATION STUDIES IN THE TEVATRON\\ WITH ELECTRON
LENSES}

\author{Giulio~Stancari\thanks{E-mail: stancari@fnal.gov.} and
  Alexander~Valishev\\
  Fermi National Accelerator Laboratory, Batavia, IL 60150, USA}

\maketitle

\begin{abstract}
  At the Fermilab Tevatron collider, we studied the feasibility of
  suppressing the antiproton head-on beam-beam tune spread using a
  magnetically confined 5-keV electron beam with Gaussian transverse
  profile overlapping with the circulating beam. When electron cooling
  of antiprotons was applied in regular Tevatron operations, the
  nonlinear head-on beam-beam effect on antiprotons was
  small. Therefore, we first focused on the operational aspects, such
  as beam alignment and stability, and on fundamental observations of
  tune shifts, tune spreads, lifetimes, and emittances. We also
  attempted two special collider stores with only 3~proton bunches
  colliding with 3 antiproton bunches, to suppress long-range forces
  and enhance head-on effects. We present here the results of this
  study and a comparison between numerical simulations and
  observations. These results contributed to the application of this
  compensation concept to RHIC at Brookhaven.
\end{abstract}

\section{INTRODUCTION}

The nonlinear forces between colliding beams are one of the main
performance limitations in modern colliders. Electron lenses have been
proposed as a tool for mitigation of beam-beam
effects~\cite{Shiltsev:PRSTAB:1999}.  It was demonstrated that the
pulsed electron current can produce different betatron tune shifts in
different proton or antiproton bunches, thus cancelling bunch-to-bunch
differences generated by long-range beam-beam
forces~\cite{Shiltsev:PRL:2007, Shiltsev:NJP:2008,
  Shiltsev:PRSTAB:2008}. In these experiments, the electron beam had a
flat transverse current-density distribution, and the beam size was
larger than the size of the circulating beam. To first order, the
effect of the electron lens was a bunch-by-bunch linear betatron tune
shift.

The present research went a step further.  We studied the feasibility
of using the magnetically confined, nonrelativistic beam in the
Tevatron electron lenses to compensate nonlinear head-on beam-beam
effects in the antiproton beam. For this purpose, the transverse
density distribution of the electron beam must mimic that of the
proton beam, so that the space charge force acting on the antiprotons
is partially canceled. The betatron phase advance between the
interaction points and the electron lens should be close to an integer
multiple of~$\pi$.

During regular Tevatron operations, both stochastic and electron
cooling were used to reduce the transverse emittance of
antiprotons. Under these conditions, antiprotons were transversely
much smaller than protons, making head-on effects essentially
linear. Intensity loss rates of antiprotons due to beam-beam were
caused by long-range interactions and rarely exceeded 5\% per
hour. While an improvement of the Tevatron performance by head-on
beam-beam compensation was not foreseen, we were interested in the
feasibility of the concept and in providing the experimental basis for
the simulation codes used in the planned application of electron
lenses to the RHIC collider at BNL~\cite{Fischer:IPAC:2012,
  Gu:IPAC:2012, Luo:PRSTAB:2012}.

\section{EXPERIMENTAL APPARATUS}

\begin{figure}[b!]
\centering
\begin{tabular}{cc}
\emph{side view} & \emph{top view} \\
\includegraphics[height=0.48\columnwidth]{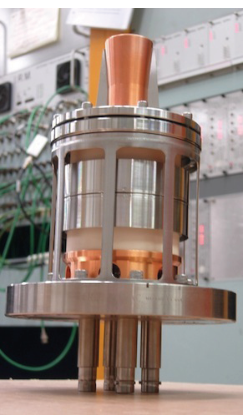} &
\includegraphics[width=0.48\columnwidth]{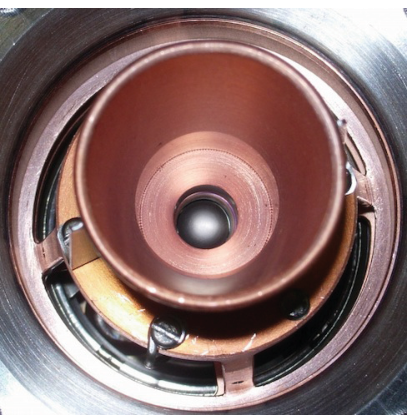} \\
\includegraphics[width=0.48\columnwidth]{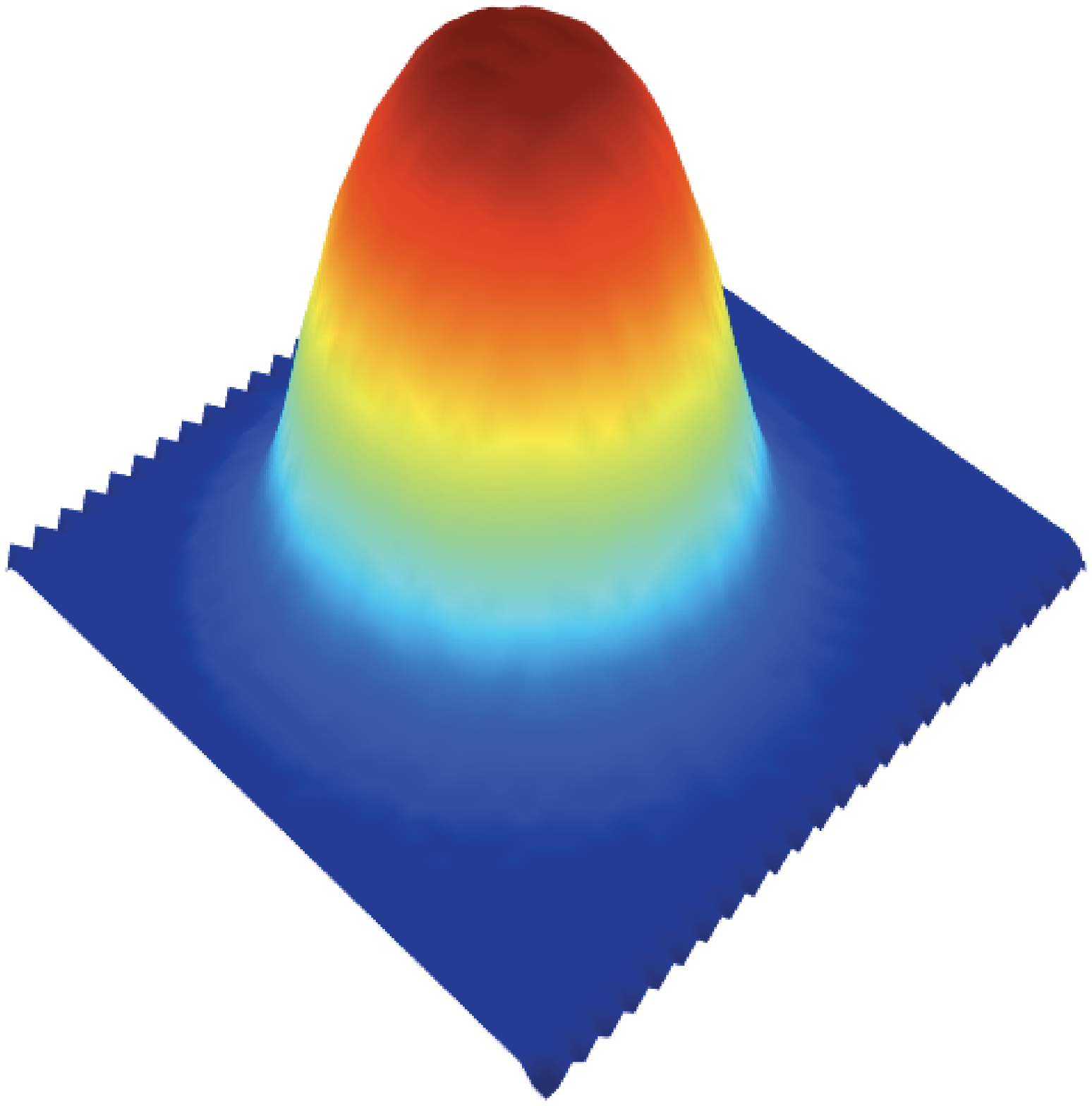} &
\includegraphics[width=0.48\columnwidth]{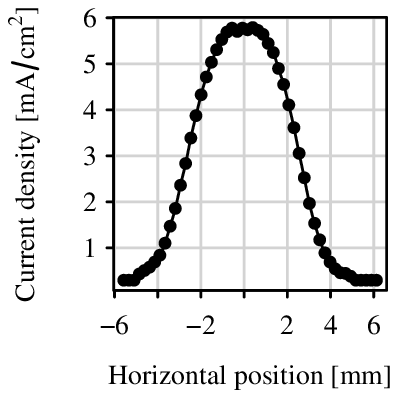} \\
\multicolumn{2}{c}{\emph{measured current-density profiles}}
\end{tabular}
\caption{The 10.2-mm (0.4-in) Gaussian electron gun: the assembled gun
  (top left); a detail of the copper cylindrical anode and of the
  convex tungsten dispenser cathode surface (top right); example of
  current-density measurements (bottom).}
\label{fig:gun}
\end{figure}

\begin{figure}[b!]
\centering
\includegraphics[width=\columnwidth]{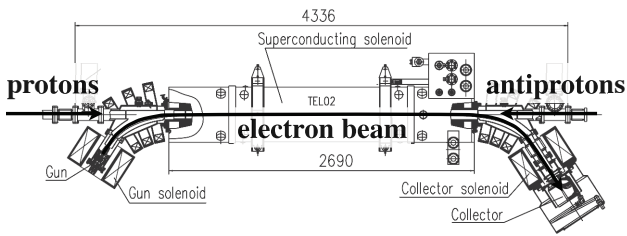}
\caption{Layout of the beams in the Tevatron electron lens. (Dimensions
  are in millimeters.)}
\label{fig:TEL}
\end{figure}

\begin{table}[t!]
  \caption{Tevatron lattice functions (amplitude~$\beta$,
    dispersion~$D$, and betatron phase~$\phi$) at the interaction points and at
    the electron lens.}
  \label{tab:lattice}
  \begin{center}
    \begin{tabular}{lrrrrrr}
      \toprule
      & $\beta_x$ & $\beta_y$ & $D_x$ & $D_y$ & $\phi_x$ & $\phi_y$ \\
      & \multicolumn{2}{c}{[m]} & \multicolumn{2}{c}{[m]} &
        \multicolumn{2}{c}{[$2\pi$]} \\
      \midrule
      CDF & 0.30 & 0.30 & 0.0 & 0.0 & 6.63 & 6.85 \\
      DZero & 0.50 & 0.50 & 0.0 & 0.0 & 13.77 & 13.85 \\
      TEL2 & 68 & 153 & 1.2 & $-$1.0 & 3.17 & 3.22 \\
      \bottomrule
    \end{tabular}
  \end{center}
\end{table}

An electron gun based on a convex tungsten dispenser cathode operating
at a temperature of 1400~K was designed and
built~\cite{Kamerdzhiev:EPAC:2008}. The diameter of the cathode was
10.2~mm (0.4~in). Its shape and the geometry of the electrodes were
chosen to produce a current density profile close to a Gaussian
distribution. Figure~\ref{fig:gun} shows pictures of the electron gun
and an example of a current density measurement. The maximum peak
current yield was 0.5~A at a cathode-anode voltage of 4.6~kV. The
standard deviation (rms) of the current profile distribution was
$\sigma_g = \q{2.0}{mm}$ at the gun.

The electron gun was installed in the second Tevatron electron lens
(TEL2) in June~2009 (Figure~\ref{fig:TEL}). In the electron lens, the
beam was generated inside the gun solenoid (0.1--0.4~T) and guided by
a superconducting solenoid (1--6~T) through the 3-m overlap region,
where it interacted with the circulating beams (protons or
antiprotons) before being extracted and dumped in the collector.  The
size~$\sigma_m$ of the electron beam in the overlap region was
controlled by the ratio between the magnetic field in the gun
solenoid~$B_g$ and in the main solenoid~$B_m$: $\sigma_m = \sigma_g
\cdot \sqrt{B_g/B_m}$. Distortions of the electron beam profile due to
its space-charge evolution were mitigated by the large axial field
($B_m > \q{1}{T}$).

In the Tevatron, 36 proton bunches (referred to as P1--P36) collided
with 36 antiproton bunches (A1--A36) at the center-of-momentum energy
of 1.96~TeV. There were 2 head-on interaction points (IPs),
corresponding to the CDF and the DZero experiments. Protons and
antiprotons circulated in the same vacuum pipe on helical
orbits. Their separation at TEL2 was 9~mm (about 6~mm both
horizontally and vertically). Each particle species was arranged in
3~trains of 12~bunches each, circulating at a revolution frequency of
47.7~kHz. The bunch spacing within a train was 396~ns, or 21 rf
buckets at 53~MHz. The bunch trains were separated by 2.6-$\mu$s
abort gaps. The synchrotron frequency was 34~Hz, or $7\times 10^{-4}$
times the revolution frequency. The machine operated with betatron
tunes near 20.58. The relevant lattice functions are reported in
Table~\ref{tab:lattice}. Thanks to the special 5-kV high-voltage
modulator (200-ns rise time), the electron beam could be synchronized
with any bunch or group of bunches, and its intensity could be varied
bunch by bunch~\cite{Pfeffer:JINST:2011}.

\section{RESULTS}

Experiments on beam-beam compensation with Gaussian electron beams
were carried out between September~2009 and July~2010. Preliminary
results were discussed in Refs.~\cite{Valishev:IPAC:2010,
  Valishev:PAC:2011}.

\subsection{Beam Alignment and Loss Patterns}

\begin{figure}
\centering
\includegraphics[width=\columnwidth]{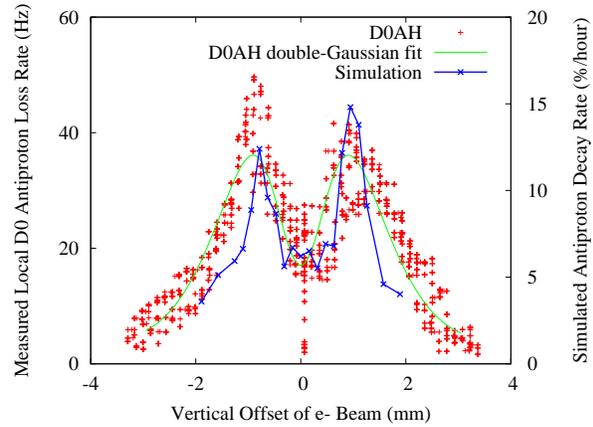}
\caption{Measured loss rates (red) and calculated intensity decay
  rates (blue) during a vertical electron beam scan across the
  antiproton beam. The antiproton vertical tune was lowered by 0.003
  to enhance the effect. No losses caused by the electron beam were
  observed with nominal tunes.}
\label{fig:vscan}
\end{figure}

Because of the nonlinear fields, alignment between electrons and
antiprotons was critical. We performed several position scans to
ensure that the response of the beam position monitors was accurate
for both fast signals from antiproton bunches and for slower signals
from electron pulses. These position scans were also useful to assess
the effects of misalignments on losses and to compare the experimental
results with numerical calculations.  We simulated losses during a
vertical alignment scan using the weak-strong numerical tracking code
Lifetrac~\cite{Shatilov:PAC:2005}. The model included the full
collision pattern for the relevant antiproton bunch and a thin-kick
Gaussian electron beam implemented via an analytical formula. The beam
parameters corresponded to the conditions at the time of the
measurement at the end of Store~7718. We tracked a bunch of 5\,000
macroparticles for $3\times 10^6$~turns for various vertical electron
beam misalignments and evaluated the intensity loss rate.  The
simulation reproduced several features observed in experiments. First,
the simulation performed at the nominal antiproton working point
(tunes set to $Q_x=0.575$, $Q_y=0.581$) predicted no losses for any
value of the vertical misalignment. This was also observed
experimentally: at the nominal working point, the electron beam did
not cause any additional beam loss. Similarly to the experiment, the
verical tune in the simulation had to be lowered by 0.003 to produce
particle losses.  Moreover, the simulation at the modified working
point demonstrated the characteristic double-hump structure of the
loss rate as a function of offset. The position of peaks was in good
agreement with the measurements. Figure~\ref{fig:vscan} shows the
measured loss rates (red crosses) and the simulated decay rates (blue
crosses and lines). Both electron and antiproton vertical rms beam
sizes in the overlap region were equal to 0.6~mm.

\subsection{Incoherent Tune Shifts and Tune Spread}

\begin{figure}
\centering
\includegraphics[width=\columnwidth]{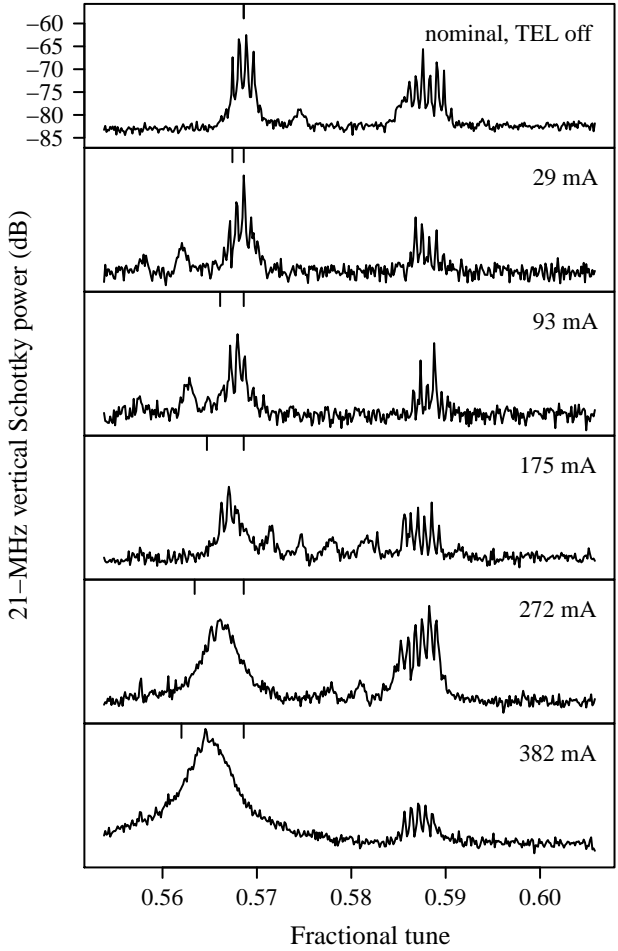}
\caption{Schottky spectra vs. electron lens current.}
\label{fig:Schottky}
\end{figure}

The effect of the electron lens on the incoherent tune distribution
could be observed directly during dedicated antiproton-only stores,
when there was no contamination from protons in the 21-MHz Schottky
signal.  Figure~\ref{fig:Schottky} shows the vertical Schottky signal
as a function of electron lens current. The vertical tick marks
indicate the expected magnitude of the linear beam-beam
parameter~$\xi_e$ due to $N_e$~electrons with Gaussian standard
deviation~$\sigma_e$ and velocity~$\beta_e c$ at a location where the
amplitude function is~$\beta$:
\begin{equation}
\xi_e = -\frac{N_e r_p \beta (1+\beta_e)}{4\pi \gamma_p \sigma_e^2}.
\end{equation}
Here, $r_p$ represents the classical radius of the proton and
$\gamma_p$ is the relativistic factor of the circulating beam. As
expected, a downward shift and widening of the antiproton tune
distribution is observed. The width of the vertical tune line agrees
well with the hypothesis that $\xi_e$ represents the maximum tune
shift.

\subsection{Effects on Coherent Beam-beam Modes}

\begin{figure}
\centering
\includegraphics[width=\columnwidth]{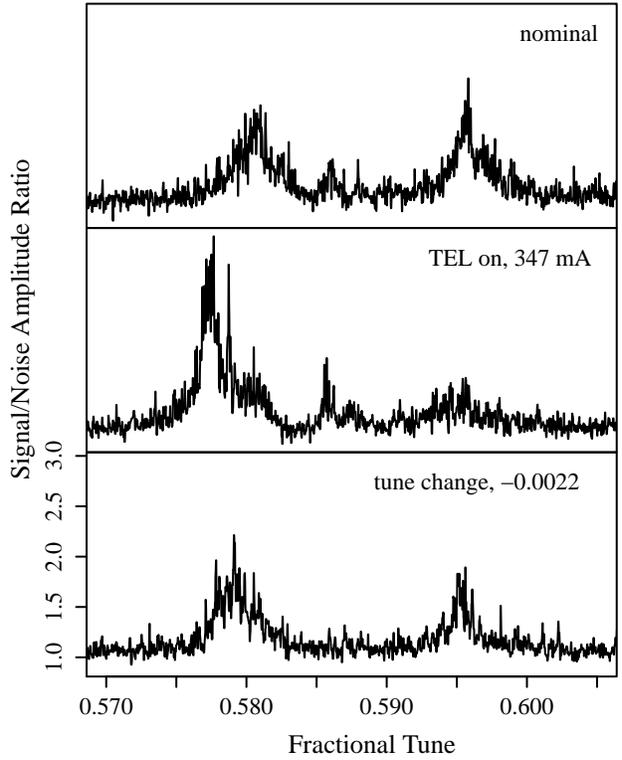}
\caption{Spectra of transverse coherent modes.}
\label{fig:tcm}
\end{figure}

A system for bunch-by-bunch measurements of transverse coherent
beam-beam oscillations was developed~\cite{Stancari:BIW:2010,
  Stancari:PRSTAB:2012}. It was based on the signal from a single
beam-position monitor in a region of the ring with high amplitude
functions. Because of its high frequency resolution and its
single-bunch capability, this system complemented the Schottky
detectors and direct-diode-detection base-band tune monitor. It was
conceived as a possible tool to monitor beam-beam compensation
effects.

Figure~\ref{fig:tcm} shows the signal from a single antiproton bunch
towards the end of a regular collider store (Store~7719). The top plot
shows the spectrum of coherent modes under nominal conditions. The
linear beam-beam parameter per interaction point was 0.0050 for
antiprotons and 0.0023 for protons. The middle plot corresponds to the
electron lens acting on the bunch, with $\xi_e = -0.006$. For
comparison, the bottom plot shows the effect of lowering the vertical
antiproton tune by~0.0022. In the middle plot, one can see a downward
shift of the first eigenmode and a suppression of the second. This
suppression could be caused in part by the antiproton tune moving away
from the proton tune. A considerable change in the width of the first
coherent mode was also observed, but relating the reduced width of the
coherent mode to a narrower tune distribution (as one would expect if
there was beam-beam compensation) requires further investigation and
numerical simulations.

\subsection{Tune Scans with Dedicated Head-on-only Stores}

\begin{figure*}
\centering
\includegraphics[width=0.9\textwidth]{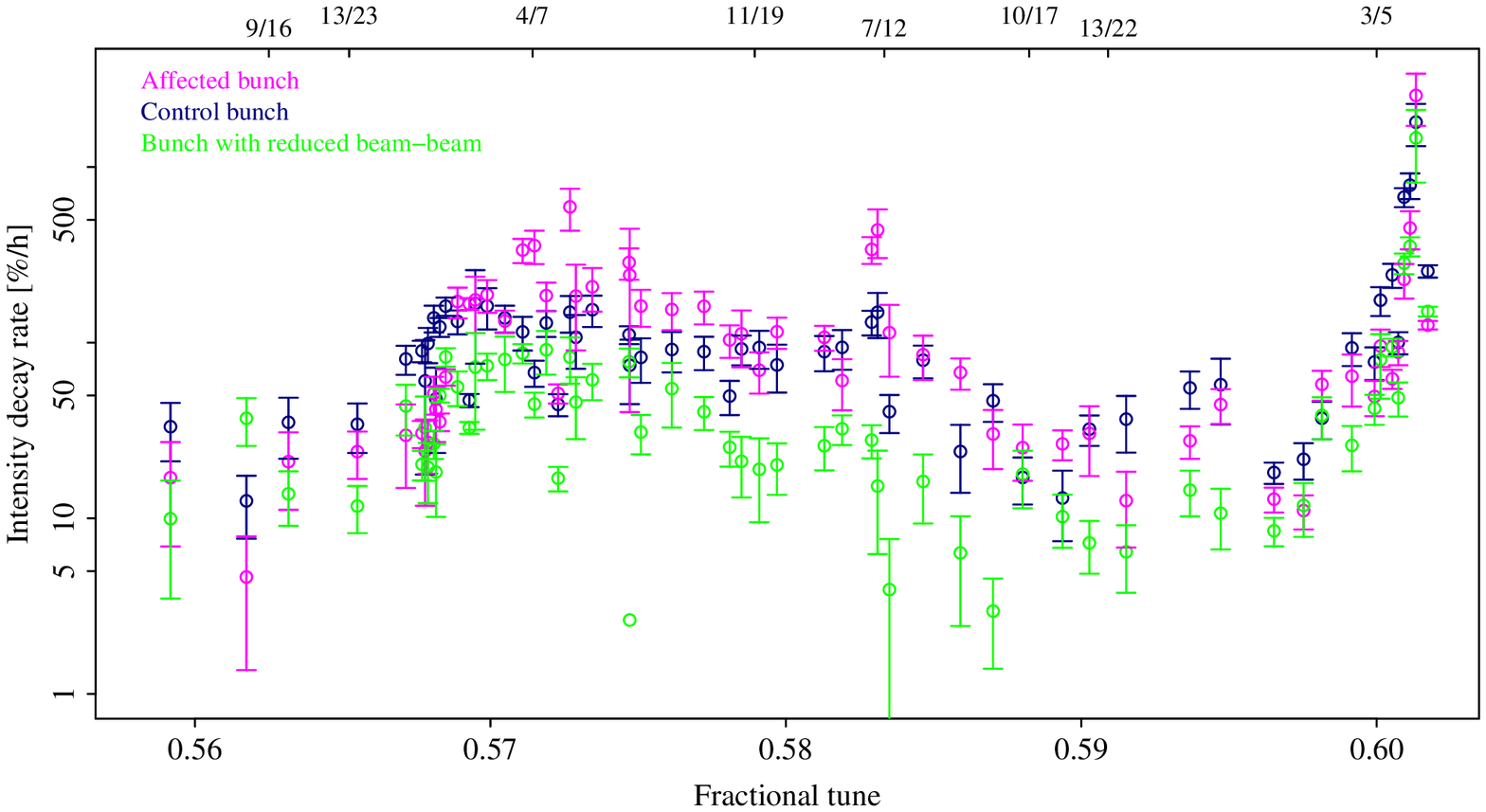}
\caption{Measured decay rates of the 3 antiproton bunches during a
  diagonal tune scan in a special 3-on-3 collider store.}
\label{fig:3x3meas}
\end{figure*}

\begin{figure}
\centering
\includegraphics[width=\columnwidth]{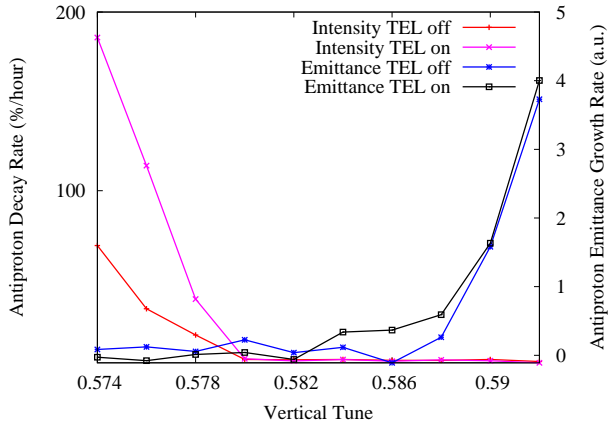}
\caption{Numerical simulation of a diagonal tune scan.}
\label{fig:3x3sim}
\end{figure}

To enhance head-on effects and to suppress long-range forces in the
Tevatron, two special 3-on-3 collider stores were attempted. In these stores,
3~proton bunches collided with 3~antiproton bunches. The bunches were
equally spaced around the machine. Antiprotons were intentionally
heated to increase their emittance and approach the size of proton
bunches.  Unfortunately, during the first experiment, the emittances
of two proton bunches increased dramatically between the beta squeeze
and collisions, before the beginning of the study. Hence, the store
could not be used for our purposes.

A smaller blow up of proton emittances occurred before the second
study as well, making conditions far from ideal: the antiproton
beam-beam parameter was less than 0.015, electron sizes could not be
matched to proton sizes, and the attempt to increase the size of the
electron beam resulted in a reduced compensation strength ($\xi_e =
-0.002$). Nevertheless, several tune scans were performed, both
vertically and diagonally in the tune diagram.

Figure~\ref{fig:3x3meas} shows the measured decay rates for the 3
antiproton bunches as a function of the average tune (from the 1.7-GHz
Schottky detector) during a diagonal scan: the bunch affected by the
electron lens (A25, magenta), the control bunch (A13, dark blue), and
the bunch colliding with the two least dense proton bunches (A1,
green). Lifetimes and tune space were obviously better for~A1. The
tune shift of the affected bunch with respect to the control bunch is
compatible with the expected amount (0.002), but it is too small to be
clearly observed. Some resonances (4/7 and 7/12, for instance) appear
stronger with the lens on, whereas the 3/5 is weaker (or shifted). One
may observe that, as expected, beam-beam forces appear to drive the
even resonance 7/12 (large difference between the green and the blue
points), but not the odd resonance 4/7 (control bunch and
low-beam-beam bunch have similar lifetimes). There are regions of the
working point where the bunch affected by the electron lens had better
lifetime (0.560--0.568 and 0.592--0.598), but this special 3-on-3
store was not enough to clearly see a reduction in tune spread or an
improvement in the available tune space.

Nevertheless, these measurements provided useful information on the
available tune space for comparisons with simulation
codes. Figure~\ref{fig:3x3sim} shows the antiproton intensity decay
rates and emittance growth rates calculated with Lifetrac as a
function of tune in a diagonal scan. The horizontal scale is the bare
lattice tune plus half of the beam-beam parameter, in order to
simulate the average of the incoherent tune distribution. As the tune
approaches the 7th order resonance (0.571) from above, loss rates
increase dramatically. Increasing the tune towards the 5th order
resonance (0.6) causes emittance growth. According to this
calculation, with the nonideal experimental conditions described
above, the electron lens does not cause harm in the stable region, but
it can make things worse outside. The region of available tune space
is well reproduced by the simulations.

\section{CONCLUSIONS}

The first studies of beam-beam compensation with Gaussian electron
lenses were carried out at the Tevatron.

We found that, in spite of the very different time structure of the
antiproton bunch and of the electron pulse, alignment of the electron
beam with the circulating beam using a common beam position monitor
was accurate to within 0.1~mm and reproducible from store to store.

We observed the effects of the electron lens on beam lifetimes and
tunes. At the nominal working point in tune space, the electron lens
did not have any adverse effects on the circulating beam, even when
intentionally misaligned. With only antiprotons in the machine, the
tune shift and tune spread caused by the electron lens were clearly
seen.

Dedicated collider stores with only 3 bunches per species (no
long-range interactions) were attempted, but the experimental
conditions were not ideal. The data was used for code
benchmarking. Moreover, tune scans conducted during these special
stores provided a direct comparison between the lifetimes of a control
antiproton bunch, a bunch affected by the electron lens, and a bunch
experiencing reduced beam-beam forces.

The machine was not ideal for a direct demonstration of the beam-beam
compensation concept for two main reasons: head-on nonlinearities for
cooled antiprotons were weak during normal operations; and the lattice
requirements (zero dispersion, phase advance close to an integer
multiple of~$\pi$) were not exactly met at the electron
lens. Nevertheless, several key experimental observations were made.

\section{ACKNOWLEDGMENTS}

The authors would like to thank W.~Fischer and C.~Montag (BNL) for
their suggestions on experiment design and for participating in part
of the studies, and V.~Shiltsev (Fermilab) for discussions and
insights. We are grateful to the Operations Department in Fermilab's
Accelerator Division for making these experiments possible.

Fermi Research Alliance, LLC operates Fermilab under Contract
No.~DE-AC02-07CH11359 with the United States Department of
Energy. This work was partially supported by the US LHC Accelerator
Research Program (LARP).

\end{document}